%% file: preprint.tex
\documentstyle[agupp,psfig]{article}
\lefthead{IZMODENOV ET AL.}
\righthead{ INTERSTELLAR ATOM FILTRATION }
\received{May 21,~1998}
\revised{November 11,~1998}
\accepted{November 11,~1998}
\paperid{1998JA90122}
\cpright{AGU}{1999}
\ccc{0148-0227/99/1998JA90122\$09.00}

\authoraddr{ V.B. Baranov and Y.G. Malama,
Institute for Problem in Mechanics,
Russian Academy of Sciences,
Prospekt Vernadskogo 101, Moscow, 117526,
Russia.
(baranov@ipmnet.ru; malama@ipmnet.ru)}

\authoraddr{ J. Geiss, International Space Science Institute,
Hallerstrasse 6, 3012 Bern, Switzerland.
(geiss@issi.unibe.ch) }
\authoraddr{ G.Gloeckler,
Department of Physics, University of Maryland, College Park, Maryland
20742. (gloeckler@umdsp.umd.edu) }
\authoraddr{ V.V. Izmodenov and R. Lallement,
Service d'Aeronomie du CNRS, BP 3, 91371 Verrieres le Buisson, France.
(izmod@ipmnet.ru; Rosine.Lallement@aerov.jussieu.fr)}

\slugcomment{To appear in the {\it Journal
of Geophysical Research}, 1999.}

\input{update.tex}
\begin{document}
\twocolumn

\title{ Filtration of interstellar hydrogen in the two-shock
heliospheric interface:
Inferences on the local interstellar electron density}

\author{
Vladislav V. Izmodenov \altaffilmark{1,2,3}, Johannes Geiss \altaffilmark{1},
Rosine Lallement \altaffilmark{2}, George Gloeckler \altaffilmark{4},
Vladimir B. Baranov \altaffilmark{5},  Yury G. Malama \altaffilmark{5}}
\altaffiltext{1}{ International Space Science Institute, Bern, Switzerland.}
\altaffiltext{2}{ Service d'Aeronomie, CNRS, Verrieres le Buisson, France.}
\altaffiltext{3}{ Permanently at Department of Aeromechanics and Gasdynamics
of Moscow State University, Mechanics and Mathematics faculty, Moscow, Russia.}
\altaffiltext{4}{ Department of Physics, University of Maryland, College Park.}
\altaffiltext{5}{ Institute for Problems in Mechanics, Russian Academy of 
Science, Moscow, Russia.}

\begin{abstract}
The solar system is moving through the partially ionized local
interstellar cloud (LIC).
The ionized matter of the LIC interacts with the expanding solar wind
forming the heliospheric interface.
The neutral component (interstellar atoms) penetrates through the
heliospheric interface into the heliosphere,
 where it is measured directly \textquotedblleft in situ" 
 as pick-up ions and neutral atoms (and as anomalous cosmic rays) 
 or indirectly through resonant scattering of solar Ly  $\alpha$.
 When crossing the heliospheric interface, interstellar atoms interact
with the plasma component through charge exchange.
 This interaction leads to changes of both neutral gas and plasma properties. 
 The heliospheric interface is also the source of radio emissions which have
  been detected by the Voyager since 1983.
 In this paper, we have used a kinetic model of the flow of the interstellar atoms 
 with updated values of velocity, temperature, and density of 
 the circumsolar interstellar hydrogen and calculated how all quantities 
 which are directly associated to the observations vary as a function of the 
 interstellar proton number density $ n_{\rm p, LIC} $. These quantities are
the degree of filtration, the temperature and the velocity of the
interstellar H atoms in the inner heliosphere, 
the distances to the bow shock (BS), heliopause, and termination shock, 
and the plasma frequencies in the LIC,
 at the BS and in the maximum compression
 region around the heliosphere which constitutes the 
\textquotedblleft barrier" for radio 
waves formed in the interstellar medium.
Comparing the model results 
with recent pickup ion data, Ly $\alpha $ 
measurements, and low-frequencies radio emissions, we have searched for
a number density of protons in the local interstellar
cloud compatible with all observations.

 We find it difficult in the frame
 of this model without interstellar magnetic field to reconcile the distance to the shock and heliopause deduced 
 from the time delay of the radio emissions with other diagnostics and
 discuss possible explanations for these discrepancies, 
 as the existence of an additional interstellar magnetic pressure  
 (2.1  $\mu$G $<$ B $<$ 4 $\mu$G  for a perpendicular magnetic field).
We also conclude that on the basis of this model the most likely
value for the proton density in the local interstellar cloud is in the range
 0.04 cm$^{-3} < n_{\rm p, LIC} < $ 0.07 cm$^{-3}$.

\end{abstract}

\begin{article}

\section{Introduction}

Our solar system is moving through a partially ionized interstellar
cloud. The ionized fraction of this local
interstellar cloud (LIC) interacts with the expanding solar wind and
forms the LIC-solar wind (SW) interface (or
heliospheric interface). The characterization of this interface is 
a timely major objective
in astrophysics and space plasma physics. The interest to the construction
of the LIC/SW interaction models is increasing at the present time
[{\it Ripken and Fahr}, 1983; {\it Baranov and Malama}, 1993;
{\it Zank et al.}, 1996; {\it Linde et al.}, 1998; 
{\it Pogorelov and Matsuda}, 1998]. 
The choice of an adequate model of the interface
depends on the parameters of the LIC. Some of these parameters, as the
Sun/LIC relative velocity, or
the LIC temperature are now well constrained [{\it Witte et al.}, 1993;
{\it Lallement and Bertin}, 1992;
{\it Linsky et al.}, 1993; {\it Lallement et al.}, 1995; see, also,
{\it Frisch}, 1995], 
but unfortunately there are
no direct ways to measure the circumsolar interstellar
electron (or proton) density, nor the local interstellar magnetic field,
while these two parameters govern the structure and the size of our
heliosphere. There have been measurements of the average electron density
in the LIC toward nearby stars.  
However, resulting densities range from 0.05 (-0.04, +0.14) cm$^{-3}$ to up to 
0.3 (-0.14, +0.3) cm$^{-3}$ depending on the ions used for the diagnostics or 
on which line-of-sight is probed  [e.g., {\it Lallement and Ferlet}, 1997]. The most
precise, temperature independent value is 0.11  cm$^{-3}$ toward the star
 Capella [{\it Wood and Linsky}, 1997]. In addition, what is measured is always
 averaged over large distances, while the ionization degree in the local
interstellar medium is very likely highly variable and out of ionization 
equilibrium [e.g., {\it Vallerga}, 1998]. 
Therefore there is a need for indirect
 observations which can bring
stringent constraints on the plasma density
and on the shape and size of the interface.  Such constraints should
help to predict when the two
Voyager spacecraft will cross the interface
 and whether or not they will be able to perform and transmit direct
observations.

Among the various types of heliospheric interface diagnostics, there are
measurements of the pick-up ions.
 Pick-up ions are formed
when interstellar neutrals, having penetrated into the heliosphere, become
ionized by
 charge exchange with the solar wind ions or photoionized. 
 The newly created ions are then convected away from the Sun
 by the solar wind.
The detection of  He$^+$ and  He$^{++}$
 provides a new determination of the neutral
 helium flow properties [{\it Gloeckler et al.}, 1997], 
which can be compared with the direct detection 
of the neutral helium [{\it Witte et al.}, 1993, 1996]. 
It is interesting to note that
the different determinations of the helium density are now in rather good
 agreement.
 With the Solar Wind Ion Composition Spectrometer (SWICS) instrument on board Ulysses, 
{\it Gloeckler et al.} [1997] found
 $n$(HeI) = 0.0153 $\pm$ 0.0018 cm$^{-3} $
with an uncertainty of the order of only 8\%, compatible with
Active Magnetospheric Particle Tracer Explorers (AMPTE) results 
of {\it Moebius} [1996] and Ulysses Interstellar Neutral-Gas  instrument (GAS) 
results of {\it Witte et al.} [1996].
As for hydrogen, the first successful detection of H$^{+}$ 
pick-up ions has been 
done with the SWICS instrument on board the Ulysses spacecraft 
[{\it Gloeckler et al.}, 1993].
From the H$^{+}$  fluxes, one can infer a new value for the neutral hydrogen 
flux in the inner heliosphere. 
Indeed, {\it Gloeckler et al.} [1997] used the pick-up ions 
data to infer the interstellar hydrogen atom number density 
in the outer heliosphere 
and found $ n(\textnormal{HI})=0.115 \pm 0.025 $ cm$^{-3} $ 
with an uncertainty of 20\%.
Note that in all these works the H and He number densities are obtained 
using the classical so-called 
\textquotedblleft hot" model of interstellar neutrals flow in the heliosphere.
The H number density obtained from pickup ion measurements
 is close to the lower range of the interval derived from optical 
resonance data revised by {\it Qu\'{e}merais et al.} [1996], 
i.e., 0.11-0.17 cm$^{-3} $.

The neutral H density in the inner heliosphere
 is dependent on the perturbations 
the neutral H suffers at the heliospheric interface.
Since interstellar helium is not depleted in the heliospheric
interface region (n(He) heliospheric $\approx$ n(He) interstellar), 
it is possible 
to compare the properties of the neutral hydrogen
and helium flows.
Taking into account the newly derived neutral hydrogen to neutral helium 
ratio in the LIC [{\it Dupuis et al.}, 1995] 
\[
 \frac{n_{\rm LIC}\textnormal{(HI)}}{n_{\rm LIC}\textnormal{(HeI)}}=14 \pm 1
\]
{\it Lallement} [1996] 
related the neutral H density in the heliosphere 
to the plasma density in the LIC,  using a proxy for the filtration ratio of H 
as a function of the plasma density. The proxy was derived from 
{\it Baranov and Malama} [1993] model filtration ratios. It was found that 
if n(H) in the inner heliosphere is of the order 
of 0.15 (0.10, respectively) cm$^{-3} $,
 then the electron density in the circumsolar medium 
is close to 0.05 (0.11, respectively) cm$^{-3} $.
{\it Gloeckler et al.} [1997], using the new SWICS pick-up ions results 
and an interstellar HI/HeI ratio of  $ 13 \pm 1 $
 (the average value of the ratio toward the nearby white dwarfs), concluded
that $ n_{\rm LIC}(\textnormal{HI}) = 0.2 \pm 0.03$ cm$^{-3} $, which corresponds to
a filtration factor (the ratio of atom number density in the outer
heliosphere to atom number density in the LIC)
\begin{equation}
\chi = \frac{n_{\rm TS}({\rm HI})}{n_{\rm LIC}({\rm HI})} = 0.58 \pm 0.15
\end{equation}
where subscript TS is termination shock. Then, on the basis of estimates of the charge-exchange processes,
they obtained for the interstellar proton (or electron) number
density $ n_{\rm p, LIC} =0.04 \pm 0.017 $ cm$^{-3} $.

For a given plasma density, there is a nonnegligible 
influence of the neutral density 
in the LIC on the filtration ratio. 
The filtration ratios used by {\it Lallement} [1996] 
were taken from a model with 
$ n_{\rm LIC}({\rm HI}) = 0.14$ cm$^{-3} $ 
[{\it Baranov and Malama}, 1993], introducing some unconsistency in the method.
 In what follows, we will make use of 
the more appropriate value  
$ n_{\rm LIC}({\rm HI}) = 0.2$ cm$^{-3}$, 
which is based on the 
well-measured neutral helium density and the interstellar ratio measured
with the Extreme Ultraviolet Explorer (EUVE).

The goal of this paper is to determine a range for
 the interstellar proton number density $n_{\rm p,LIC}$
that is compatible with all observations, 
using the two-shock heliospheric interface model of
{\it Baranov and Malama} [1993, 1995, 1996] for the updated value
$ n_{\rm LIC}({\rm HI}) = 0.2$ cm$^{-3}$.
The observations we will consider are, in addition to the pick-up ions quoted above, the temperature and velocity of the neutral H flow in the inner heliosphere and the Voyager radio emissions.
Recently, {\it Linsky and Wood} [1996] have detected the heated and decelerated
gas from the so-called H wall corresponding to the compressed region
between the bow shock (BS) and the heliopause (HP), in absorption toward the star $\alpha$ Centauri.
{\it Gayley et al.} [1997] have compared the observed absorption with 
the theoretical absorption for three different models. 
One of these models is a two-shock model, and the two other correspond to the 
\textquotedblleft subsonic" case; That is, they have modified the equation of state of the gas to simulate the effect of an interstellar magnetic field. 
These authors conclude that the H wall absorption favors the 
\textquotedblleft subsonic case." 
We have not included such diagnostic in our study 
for the following reasons: 
in our \textquotedblleft supersonic" case, the simulations show
that it is hard to distinguish 
 H walls built up for $ n_{\rm p, LIC}=0.04 $ cm$^{-3}  $ or for  0.2 cm$^{-3}  $.
As a matter of fact, if $ n_{\rm p,LIC}$ increases, the gas is more heated and 
compressed,
but the thickness of the H wall is reduced. 
Also, the precision required to model the differences 
between the theoretical absorptions, 
namely small differences of the order of a few kilometers per second  
at the bottom of the lines, is of the order of the differences 
between the model results from different groups 
for the same parameters in the supersonic case 
[see {\it Williams et al.}, 1997, Appendix B].  
A larger difference may exist between
the \textquotedblleft supersonic" and the \textquotedblleft subsonic" cases, large enough to favor the subsonic case, as argued by Gayley et al. Our approach is to use other independent diagnostics, namely all heliospheric data, in order to constrain the requirements for additional physics.   
\vspace{0.5cm}
\begin{figure}[!t] \label{fig1}
\psfig{figure=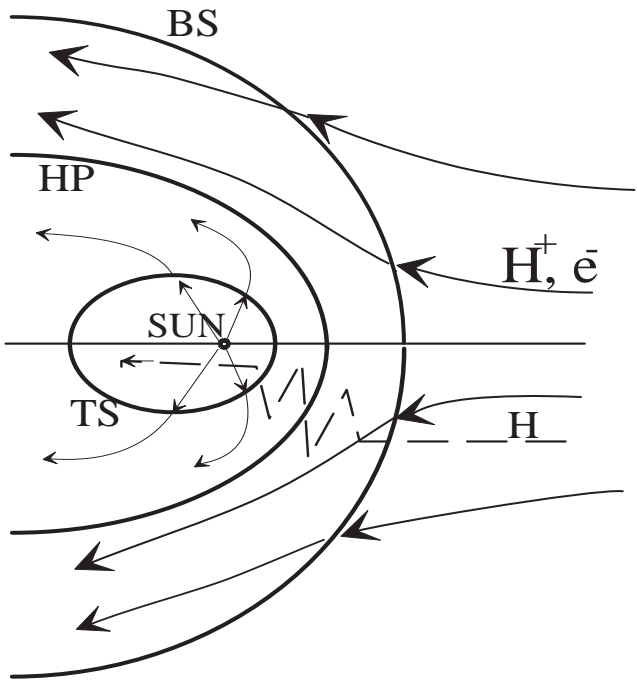,height=9.0cm,width=8.4cm,angle=0}
\caption{\scriptsize
Qualitative picture of the solar wind interaction with \\
the local interstellar medium (LISM).  
Here BS is the bow shock, \\ 
HP is the heliopause, 
and TS is the termination shock.}
\end{figure}
\vspace{0.5cm}

The model is a self-consistent
gasdynamic-kinetic model of the solar wind 
interaction with the local interstellar
medium, which takes into account the mutual influence of the plasma
component of the LIC and interstellar H atoms
in the approximation of axial symmetry (Figure 1).

Then we compare 
the range of $ n_{\rm p,LIC} $ values derived from the 
pick-up measurements with the range that is compatible with
observations of the backscattered  solar Ly $\alpha$ emission 
as well as from the interpretation of the 2-3 kHz emission
recorded by the Voyager radio 
instruments.
As a tool for the analysis of future measurements, 
we also calculate the relevant observational 
parameters as a function 
of the interstellar plasma density.

\section{Formulation of the Problem}

The interface is characterized by three surfaces: 
the solar wind termination shock (TS), 
the heliopause, and the interstellar bow shock. 
The interstellar atom flow in the heliospheric interface must be described
kinetically because the mean free path of the neutral atoms is of the
order of the size of the heliospheric interface. 
In fact, {\it Baranov et al.} [1998] have shown
kinetic and hydrodynamic models of the interstellar atom flow
may give significant differences. 
In order to obtain the kinetic distribution function,
the Boltzman equation must be solved:
\begin{eqnarray}
\vec{w_{\rm H}} 
\frac{\partial f_{\rm H}(\vec{r}, \vec{w}_{\rm H}) }{\partial \vec{r} }
+ \frac{F}{m_{\rm H}} 
\frac{\partial f_{\rm H} (\vec{r}, \vec{w}_{\rm H})}{\partial
\vec{w}_{\rm H} } = \nonumber \\
- f_{\rm H} (\vec{r}, \vec{w}_{\rm H}) 
\int | \vec{w}_{\rm H} - \vec{w}_{\rm p} |
\sigma^{\rm HP}_{ex} f_{\rm p}
(\vec{r}, \vec{w}_{\rm p}) d \vec{w}_{\rm p}   \\
+ f_{\rm p} (\vec{r}, \vec{w_{\rm H}}) \int | \vec{w}_{\rm H^*} - \vec{w}_{\rm H} |
\sigma^{\rm HP}_{ex} f_{\rm p}
(\vec{r}, \vec{w}_{\rm H^*} ) d \vec{w}_{\rm H^*}  \nonumber \\
- ( \beta_i + \beta_{\rm impact} ) f_{\rm H} ( \vec{r}, \vec{w}_{\rm H} ) \nonumber
\end{eqnarray}
Here $ f_{\rm H} ( \vec{r}, \vec{w}_{\rm H} ) $ is distribution function
of the H atoms,
$ f_{\rm p}( \vec{r}, \vec{w}_{\rm p}) $ is the local distribution
function of the protons which is assumed to be Maxwellian,
$ \vec{w}_{\rm p} $ and $ \vec{w}_{\rm H} $ are the individual proton
and H atom velocities, respectively.
Here $ \sigma^{\rm HP}_{ex} $ is the charge exchange cross section 
of an H atom with a proton,
$ \beta_i $ is the photoionization rate, $ m_{\rm H} $ is the mass of atom,
$ \beta_{\rm impact} $ is the electron impact ionization rate,
and $ F $ is the sum of solar gravitational force
and solar radiation pressure force.

Equation (2) takes into account the following processes.

1.The resonance charge exchange process:

\[
\textnormal{H + H}^+ \leftrightarrow \textnormal{H}^+ + \textnormal{H} 
\]
with charge exchange cross section [{\it Maher and Tinsley}, 1977] 
$ \sigma^{\rm HP}_{ex} = 10^{-14}$ (1.64 - 0.0695 ln V)$^2$, 
cm$^2$.
Here V is the relative velocity measured in centimeters per second.

2. The photoionization process: The photoionization rate is
\[
\beta^{\rm H}_i= \beta^{\rm H}_{ph,E} \lgroup \frac{r_e}{r} \rgroup ^2, 
\beta^{\rm H}_{ph,E}
= 8.8 \cdot 10^{-8} \: \textnormal{s}^{-1}
\]
where $ \beta^{\rm H}_{ph,E} $ is the photoionization rate at the Earth's orbit
and $ r_e $ is 1 AU.

3. The electron impact ionization process: The rate is given by [{\it Lotz}, 1967]
\begin{eqnarray}
\beta_e = \frac{2}{P_1} \sqrt{ \frac{2}{P_1 m_e \pi} }  a 
\sqrt{\lambda } \{
  E_1( \lambda)  \\
 - b \cdot e^c \frac{\lambda}{\lambda +c}
E_1(\lambda +c) \} \nonumber
\end{eqnarray}
Here $ \lambda = \frac{P_1}{k T_e}$, $ E_1 ( \lambda ) =
\int^{\infty}_1\frac{ e^(- \lambda t)}{t} d t $.
The values $  P_1, a,b,c $ are equal to  $ P_1=13.6$ eV,
$ a= 4. \cdot 10^{-14}$ cm$^2$ eV$^2$, b=0.6, c=0.56. 

4. The solar gravitation ($ F_g $) and solar radiation pressure ( $ F_r $ )processes:
\[
F = F_g - F_r = ( 1 - \mu)F_g, \mu = F_r/F_g
\]
We have used the value $ \mu = 0.8 $. Most of our results are independent 
of the chosen value, since $ F$ has a nonnegligible influence only 
within a few AU from the Sun.

The difficulty with modeling the H atom flow lies 
in the necessity to take into account the mutual 
influence of the atomic and plasma components
[{\it Baranov and Malama}, 1993, 1995, 1996] 
and to solve the kinetic equation (2)
together with hydrodynamic equations
for the plasma component. 
To calculate the H atom flow, we used the axisymmetric
model and the method
developed by {\it Malama} [1991] 
and {\it Baranov and Malama} [1993, 1995, 1996].
The boundary conditions for the proton density, the bulk velocity and
the Mach number of the
solar wind at the Earth's orbit are taken as $ n_{{\rm p}, E}=7.00$ cm$^{-3}$, 
V$_E= 450$ km s$^{-1}$, M$_E$=10.

In the unperturbed LIC, we use $ V_{\rm LIC} = 25$ km s$^{-1}$ and $ T_{\rm LIC}
= 5600$ K  for all sets of model parameters. 
These values are close to the most recent determinations
of interstellar He parameters obtained 
by {\it Witte et al.} [1996] with the GAS instrument on Ulysses.
These authors give  an interstellar helium velocity 
$ 24.6 \pm 1.1 $ km s$^{-1}$
and a helium temperature of $ 5800 \pm 700 $ K. 
The H atom number density is kept fixed at 
$ n_{\rm H, LIC}= 0.2$ cm$^{-3} $, as discussed in the introduction.
This value corresponds to the mean value given
 by {\it Gloeckler et al.} [1997].
For our calculations we have chosen the following
values of proton number density:
$ n_{\rm p, LIC}= 0.3, 0.2, 0.1, 0.07, 0.04$  cm$^{-3} $.

\section{ Results of Modeling Calculations}

Using the heliospheric interface model with the solar wind and LIC
parameters described above, we have calculated the structure 
of the heliospheric interface (positions and shapes of the TS, HP, and BS) 
and the  distributions of plasma and neutral components.
The distributions of the plasma and of the different H atom populations 
as well as the influence of the different physical processes
have been discussed  by {\it Baranov and Malama} [1993, 1995, 1996]
and {\it Baranov et al.} [1998].
Here we  present and discuss only selected results of our calculations,
which will be useful
for analyses of interstellar proton number density in the section 4.
Figure 2 shows the BS, HP,and TS distances to the Sun in the upwind direction 
as a function of the interstellar proton number density.
It can be seen from the figure that the BS has the largest response to
the interstellar proton density variations.
As a matter of fact, the distance to the BS in the upwind direction decreases
from 360 AU (for $n_{\rm p, LIC}=0.04$ cm$^{-3} $) to 180 AU 
(for $ n_{\rm p, LIC}=0.3$ cm$^{-3} $), whereas 
the distance to the heliopause varies from 185 to 110 AU. 
Thus the heliospheric interface region (the region between the BS and TS)
becomes narrower while the interstellar proton number density is increased.
The distance to the termination shock is NOT very sensitive 
to the proton density (between $ \approx$ 100 and $ \approx$ 70 AU). 
The inferred range is compatible with the shock location deduced 
from the radial gradients of the ACR energy spectra 
[e.g. {\it Cummings and Stone}, 1996].

\begin{figure}[!t] \label{fig2}
\psfig{figure=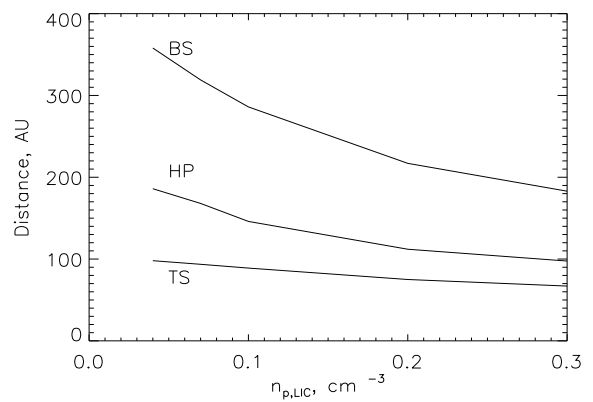,height=5.8cm,width=8.4cm,angle=0}
\caption{\scriptsize
Positions of the bow shock, the heliopause and the \\ 
termination shock
in upwind direction as a function of interstellar  \\ proton number density.}
\end{figure}
\begin{figure} \label{fig3}
\psfig{figure=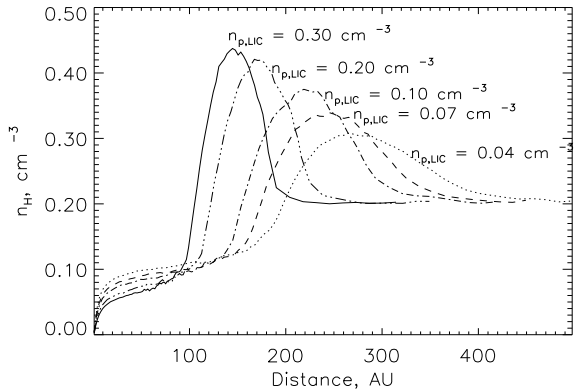,height=5.8cm,width=8.4cm,angle=0}
\caption{\scriptsize
The number density of interstellar atoms
as a function \\ 
of heliospheric distance for different values of
interstellar proton \\ number density.
}
\end{figure}
\vspace{0.5cm}
Figure 3 shows the number density of interstellar H atoms
as a function of the heliocentric distance in the upwind direction.
As the proton number density increases, 
the hydrogen wall between the BS and the HP becomes
\textquotedblleft higher." 
The filtration factor (defined in equation (1)) decreases.
Indeed, an increase  
of the proton number  density in the LIC leads to
increase of the proton number density
between the BS and the HP and 
to an increase of the secondary H atoms number density
resulting from the charge-exchange between primary interstellar
H atoms and decelerated protons.
This population of secondary H atoms has a smaller bulk velocity
and a higher temperature than
the primary interstellar H atom population.
It is the increase of the number density of the secondary H atoms which 
reinforces the H wall and decreases the filtration factor.
Figure 3 also shows that filtration occurs mainly 
in the region between the BS and the HP.

\callout{Table 1} shows how the number density of primary 
interstellar and secondary H atoms at the termination shock 
in the upwind direction changes as a function
of the density $ n_{\rm p,LIC} $.
 For $ n_{\rm p, LIC}=0.04$ cm$^{-3} $, primary 
interstellar H atoms represent about 50\% of the H atoms entering 
the heliospheric interface,
while for $ n_{\rm p, LIC}= 0.2$ cm$^{-3}$, 
they represent only 6\% of the total.
Since the secondary H atoms have a smaller velocity and a higher
temperature than the primary interstellar atoms 
[cf., {\it Baranov et al.}, 1998],
 the bulk velocity and the temperature of the mixed H atom gas
vary with the interstellar proton number density (Table 1). For example,
for $ n_{\rm p,LIC} =0.04$ cm$^{-3} $, 
the temperature and the velocity are
 $ T_{\rm H,TS} = 10500$ K  and $ V _{\rm H,TS}= 22.5$ km s$^{-1}$,
 while for $ n_{\rm p, LIC}=0.3$ cm$^{-3} $, the values are
 $ T_{\rm H,TS} = 14000$ K and 
 $ V_{\rm H,TS}= 17.0$ km s$^{-1}$.
\vspace{0.5cm}

\begin{figure}[!t] \label{fig4a}
\psfig{figure=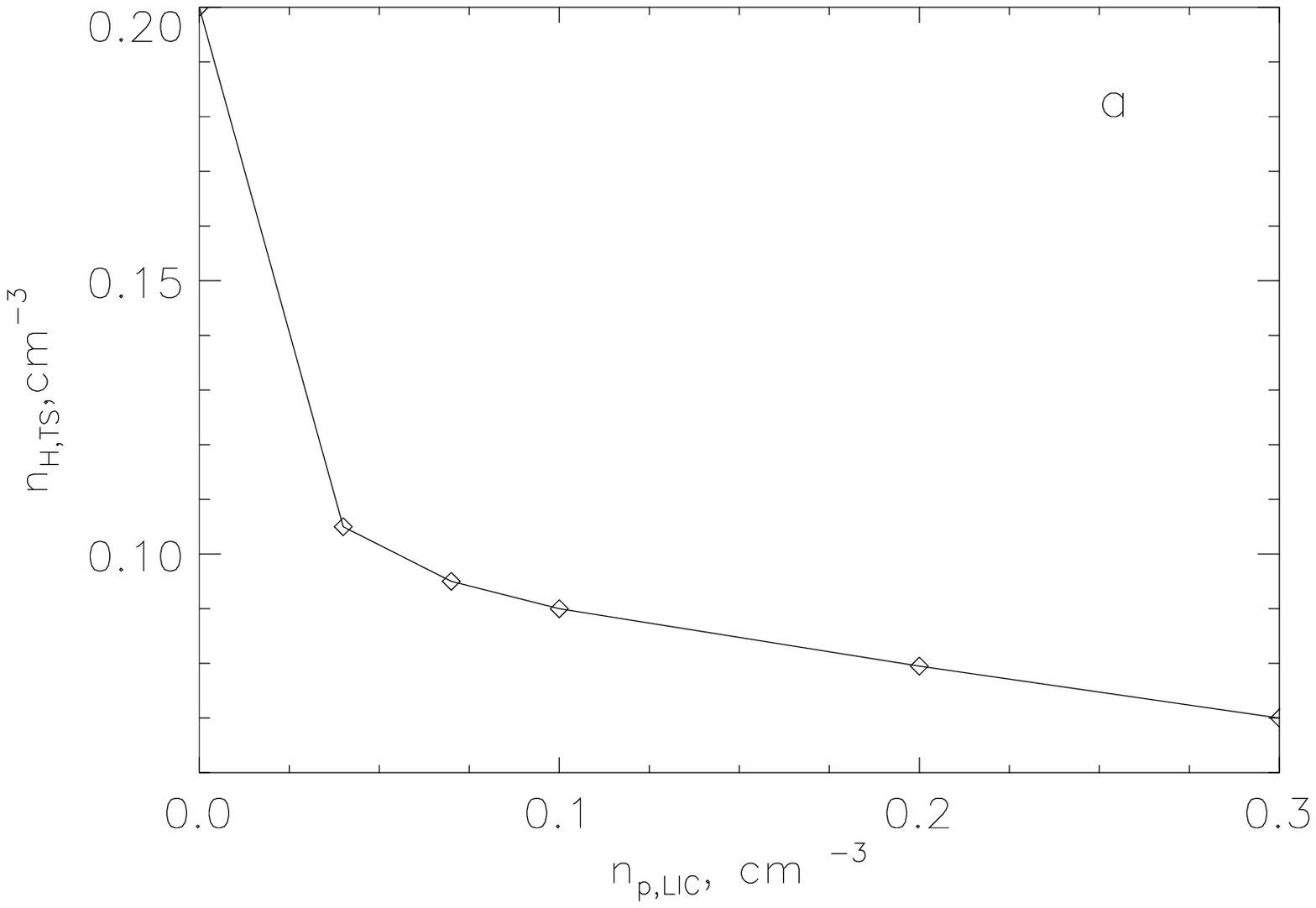,height=5.8cm,width=8.4cm,angle=0}
\end{figure}
\begin{figure} \label{fig4b}
\psfig{figure=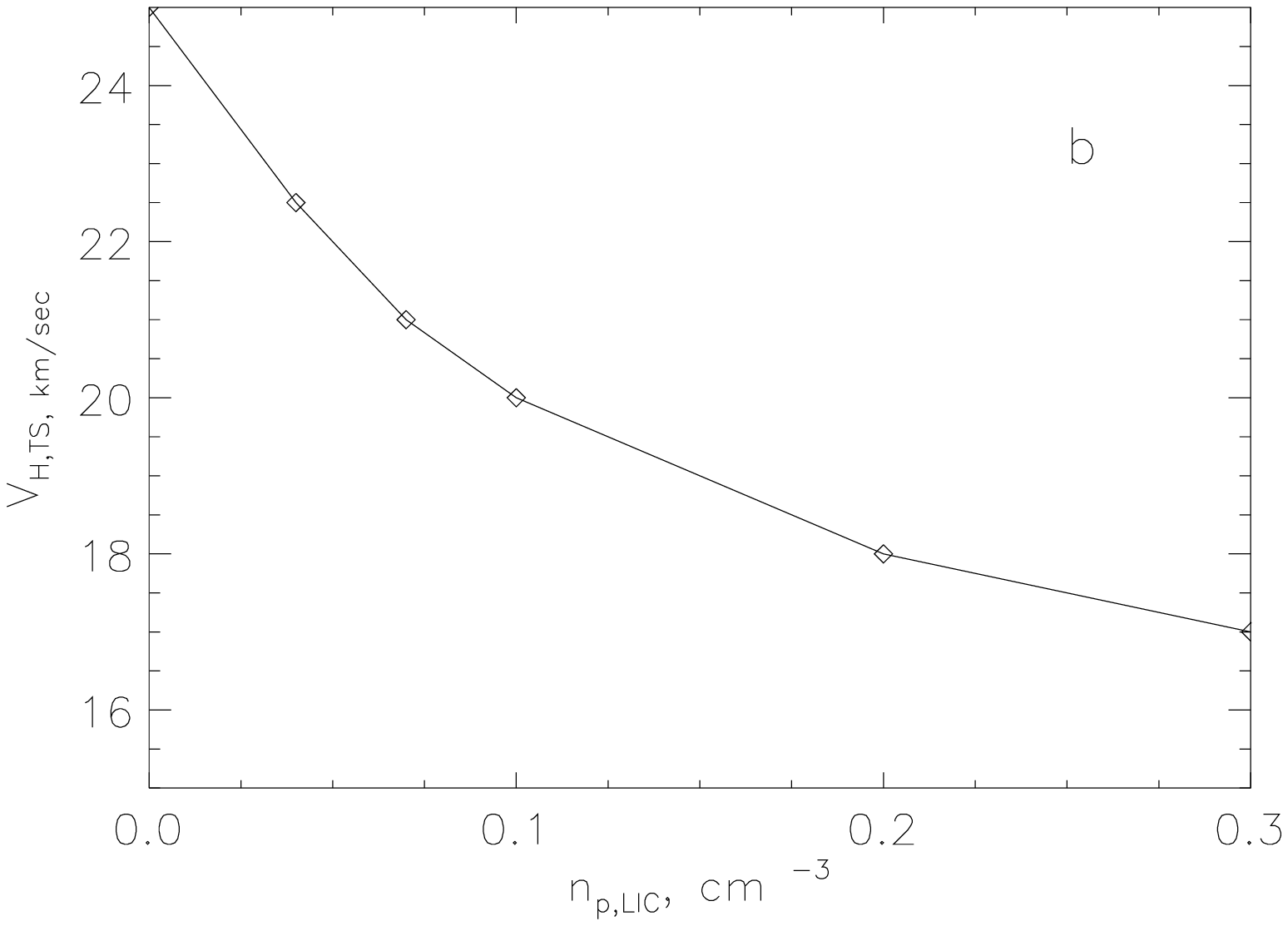,height=5.8cm,width=8.4cm,angle=0}
\end{figure}
\begin{figure} \label{fig4c}
\psfig{figure=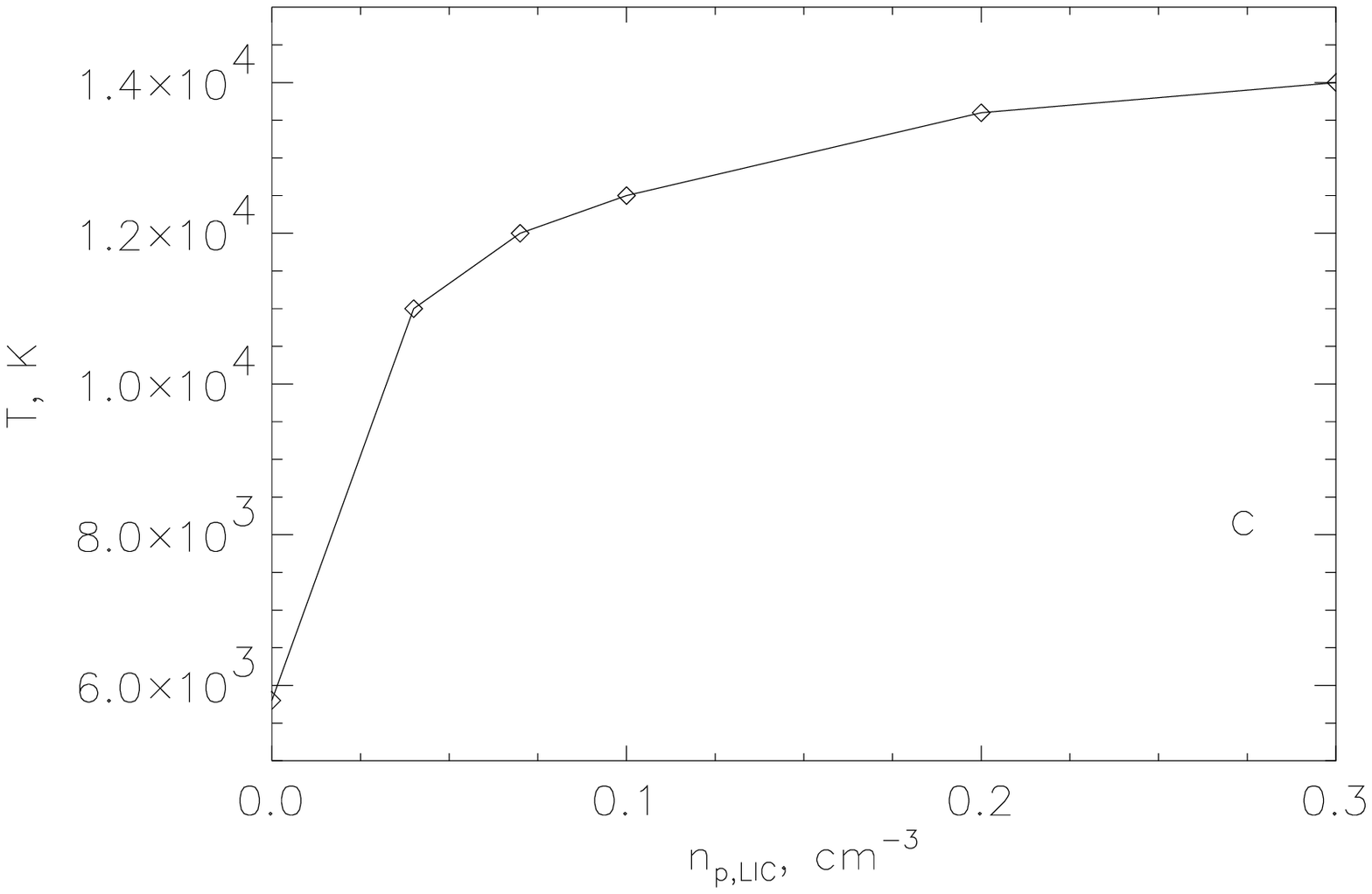,height=5.8cm,width=8.4cm,angle=0}
\caption{\scriptsize
(a) Number density, (b) the velocity, and (c) the tem- \\ 
perature of intertellar atoms
at the termination shock in upwind di - \\
rection  
as functions of  
interstellar proton number density. 
}
\end{figure}
\vspace{0.5cm}

Figures 4 shows the  interstellar atom density, velocity, and temperature at
the termination shock in the upwind direction 
as a function of $ n_{\rm p, LIC} $.
It can be seen from the figure that the interstellar atom
number density and the filtration factor decrease
 rapidly when the proton number density 
is increased from 0 to 0.04 cm$^{-3}$ and 
much less rapidly for higher $n_{\rm p,LIC} $ values.
As a consequence, small uncertainties in the atom density 
at the TS will give us
small  uncertainties in $ n_{\rm p, LIC} $ for $ n_{\rm p, LIC} < 0.04$ cm$^{-3} $
(\textquotedblleft low density case") and relatively large ones for, 
say, $n_{\rm p, LIC} > 0.07$ cm$^{-3} $ 
(\textquotedblleft high density case").
The same situation occurs for the temperature (Figure 4c), showing that its 
determination is an excellent diagnostic in the low density case.
Contrary the filtration factor and temperature, 
the velocity dependence on $ n_{\rm p, LIC} $ (Figure 4b) 
remains about the same for the full range 
 0 cm$^{-3} < n_{\rm p,LIC} < 0.3$ cm$^{-3} $,
showing that knowledge of $ V_{\rm H,TS} $ is very helpful
for determining $ n_{\rm p, LIC} $ 
even in the \textquotedblleft high density case" 
(as long as $ n_{\rm p, LIC} < 0.2$ cm$^{-3} $).

Figure 5 shows the plasma frequency in the interstellar plasma 
and at the bow shock 
on the upwind axis as a function of the proton density. 
These frequencies may be important if, 
as it has been suggested by {\it Gurnett et al.} [1993],
the 1.8 kHz emission cutoff corresponds to the interstellar plasma frequency, and if,
as suggested by {\it Grzedzielski and Lallement} [1996], the 2 kHz band was emitted 
in the compression region ahead of the bow shock.  
Also plotted is the frequency corresponding 
to the maximum plasma density along the 
upwind and crosswind axis (the \textquotedblleft pile-up" region), which according 
to the present model occurs  between the BS and the HP. 
This may be an important 
parameter too, since, 
according to the above scenario this is the \textquotedblleft obstacle" 
the 2 kHz signal has to overcome to be able to enter the heliosphere.

Recently, {\it Linsky and Wood} [1996] have convincingly shown that the excess of neutral hydrogen absorption seen in the spectrum of the star $\alpha$ Centauri had indeed its origin in the heated and decelerated
gas from the so-called H wall corresponding to the compressed region 
between the BS and the HP.
While being an important discovery, simulations show
 it is hard to distinguish from these
observations between the H walls built up for 
$ n_{\rm p,LIC}=0.04 $ or  0.2 cm$^{-3}$.
 As a matter of fact , if $ n_{\rm p, LIC}$ increases, the gas is more heated and 
compressed,
but the thickness of the H wall is reduced. 
This is why we have not included the H wall absorption in this parametric study.
A somewhat larger difference does exist between
the \textquotedblleft supersonic" and \textquotedblleft subsonic" cases, 
as calculated by {\it Gayley et al.}
Here we consider the supersonic case only.
\begin{figure} \label{fig5}
\psfig{figure=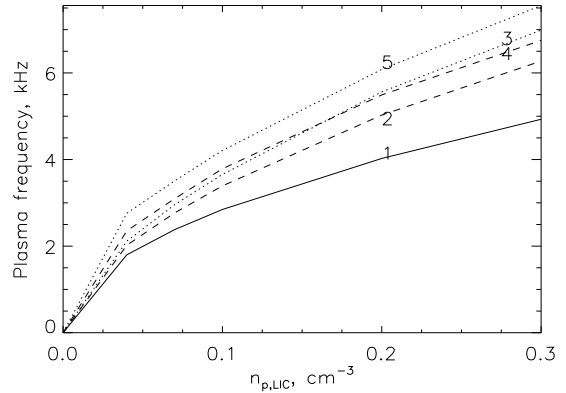,height=5.8cm,width=8.4cm,angle=0}
\caption{\scriptsize
The plasma frequencies in the
interstellar plasma  \\ (solid curve 1), at the bow shock
(curves 2 and 3), and in the ma- \\ 
ximum plasma density region between  BS and HP (curves 4 and 5) \\ 
as functions of  interstellar proton number density. 
Dotted curves  \\ 3 and 5 correspond to upwind, dashed curves 2 and 4
correspond  \\ to crosswind.
}

\end{figure}

\section{The Influence of the Interstellar Proton Number Density}

\subsection{The Neutral Hydrogen Density in the Inner Heliosphere}

The pick-up ions measurements provide a determination of the neutral H flux 
which has the advantage of being dependent on the solar ionization processes only. 
The situation is even better since the solar wind measured at the same time 
as each set of pick-ups has indeed been the main ionizing agent 
of these particular pick-ups. At variance with the pick-ups, 
the use of the backscattered Ly $\alpha$ glow 
as a neutral H density diagnostic 
suffers from uncertainties on photon radiative transfer effects 
and on radiation pressure and ionization rate measurements 
[see, e.g., {\it Qu\'{e}merais et al.}, 1994].

As mentioned above, {\it Gloeckler et al.} [1997] have obtained from
the SWICS Ulysses pick-up ion measurements for the
neutral hydrogen number density at the termination shock 
\begin{equation}
 0.09 \: \textnormal{cm}^{-3} < n_{\rm H,TS} < 0.14 \: \textnormal{cm}^{-3}
\end{equation}
using the classical \textquotedblleft hot" model.
Because the \textquotedblleft hot" model does not take 
into account any heliospheric 
filtering at the interface,
it is reasonable to assume that this measured value corresponds 
to interstellar atom number density at the TS.
Comparing this with the results of our numerical 
calculations (Figure 4a), we obtain 
\begin{equation}
0.02  \: \textnormal{cm}^{-3} < n_{\rm p, LIC} < 
0.1 \: \textnormal{cm}^{-3}
\end{equation}
\begin{planotable}{crrrrrrrrrrr}
\tablecaption{Interstellar Atoms Number Density at the Termination Shock
($n_{\rm H,LIC}=0.2$ cm$^{-3})$}
\tablenum{1}
\tablehead{
     \colhead{$n_{\rm p,LIC}$, cm$^{-3}$ }        &
     \colhead{$n_{\rm prim, TS}$, cm$^{-3} $}    &
     \colhead{$n_{\rm sec,TS}$, cm$^{-3} $}    &
     \colhead{$n_{\rm H, TS}$,  cm$^{-3} $}           &
     \colhead{$V_{\rm H, TS}$, km s$^{-1} $}               &
     \colhead{$T_{\rm H, TS}$, K }
}

\startdata
0.3   &       &       & 0.07   & 17   & 14000  \nl
0.2   &0.0045 & 0.075 & 0.0795 & 18   & 13500  \nl
0.1   &0.02   & 0.07  & 0.09   & 20   & 12500  \nl
0.07  &0.03   & 0.065 & 0.095  & 21   & 12000  \nl
0.04  &0.055  & 0.05  & 0.105  & 22.5 & 10500  \nl

\end{planotable}
\begin{planotable}{llll}
\tablewidth{41pc}
\tablecaption{Intervals of Possible Interstellar Proton Number
Densities}
\tablenum{2}
\tablehead{
     \colhead{Type of Heliospheric Interface Diagnostics }        &
     \colhead{Range of Interstellar Proton Number Density}   
}

\startdata
SWICS/Ulysses pick-up ion               & \nl  
\hspace{4pt} 0.09 cm$^{-3} < n_{\rm H,TS} < 0.14$ cm$^{-3}$  &  0.02 cm$^{-3}< n_{\rm p, LIC} < 0.1$ cm$^{-3} $  \nl  
\vspace{3pt}
\hspace{4pt} {\it Gloeckler et al.}, [1997]                  & \nl  
Ly- $\alpha$,  intensity                  & \nl   
\hspace{4pt}  0.11 cm$^{-3} < n_{H, TS} < 0.17$ cm$^{-3}$ & $ n_{\rm p, LIC} < 0.04$ cm$^{-3} $ or \nl   
\vspace{3pt}
\hspace{4pt} {\it Qu\'{e}merais et al.}, [1994]            & $ n_{\rm p, LIC} < 0.07$ cm$^{-3}$  (for $n_{\rm H, LIC}=0.23$ cm$^{-3}$ ) \nl  
Ly-$\alpha$, Doppler shift               &                 \nl   
\hspace{4pt} 18 km s$^{-1} < V_{H, TS} < 21$ km s$^{-1} $    &  0.07 cm$^{-3} <n_{\rm p, LIC} < 0.2$ cm$^{-3} $    \nl  
\hspace{4pt} {\it Bertaux et al.} [1985], {\it Lallement et al.},                    &   \nl
\vspace{3pt}
\hspace{4pt} [1996], {\it Clarke et al.} [1998] &   \nl
Voyager kHz emission (events)      &   \nl 
\hspace{4pt}  110 AU $< R_{\rm AU} <$ 160 AU               &  0.08 cm$^{-3} < n_{\rm p, LIC} < 0.22$ cm$^{-3} $  \nl  
\vspace{3pt}
\hspace{4pt} {\it Gurnett and Kurth} [1996]                 &                         \nl
Voyager  kHz emission (cutoff)      &   \nl 
\hspace{5pt}    1.8 kHz                        & $  n_{\rm p, LIC} = 0.04$ cm$^{-3}$ \nl
\hspace{5pt} {\it Gurnett et al.} [1993], {\it Grzedzielski and} & \nl
\hspace{5pt} {\it Lallement} [1996]          &  \nl 
\end{planotable}   
\clearpage
\noindent See, also, table 2.
The lower value of the neutral density interval 
falls within the 
insensitive part (\textquotedblleft high density case")
 of the function $ n_{\rm H,TS}(n_{\rm p, LIC}) $ 
in figure 4a, and this explains
the large range resulting for $n_{\rm p, LIC} $. 
This is something only estimates as those from {\it Gloeckler et al.} [1997] 
do not show and the present work makes visible.
The mean Gloeckler et al. value 
$n_{\rm H,TS}=0.115$ cm$^{-3} $ 
($ n_{\rm H,TS} / n_{\rm H,LIC} = 0.575 $)
corresponds here to $ n_{\rm p, LIC} =0.03$ cm$^{-3} $. 
This value is slightly lower than the one obtained by Gloeckler et al. 
In order to narrow the range for the interstellar proton density
on the basis of pick-up measurements,
we need a larger precision on the neutral H density.
As we mentioned above, the backscattered solar Ly  $ \alpha $ intensity can
also provide an estimate of the interstellar atom number density.
A compilation and reinterpretation
of many previous determinations of the H atom density by Ly  $ \alpha $
measurements has been done by {\it Qu\'{e}merais et al.} [1994].
The resulting density is in the range of  0.11 cm$^{-3}$ - 0.17 cm$^{-3}$, 
corresponding to 
\begin{equation}
 n_{\rm p, LIC} < 0.04 \: \textnormal{cm}^{-3}
\end{equation}
The range obtained by 
Qu\'{e}merais et al. includes the mean value
of the  $ n_{\rm p, LIC} $ determination by Gloeckler et al. 

\subsection{Velocity and Temperature}

Figure 4 shows that the temperature
and the velocity of the H atoms in the heliosphere
depend rather strongly on the LIC proton number density.
These model predictions can be compared with estimations
obtained from measurements of the backscattered
solar Ly $\alpha$ radiation. Velocities are deduced 
from Doppler shifts, while temperatures are deduced from linewidths
with the help of flow modeling. However, one has to overcome
two difficulties: \\ 
(1) The velocity determination suffers from uncertainties
in the radiation pressure. Depending on the balance between radiation 
pressure and gravitation, the velocity of the gas close to the Sun increases 
($ \mu < 1 $ valid for low activity) or decreases
($ \mu > 1 $ valid for high activity). 
Therefore an uncertainty in the balance parameter 
$\mu$  introduces an error in the H velocity far from the Sun. 
(2) The temperature determination suffers 
from uncertainties on the radiative transfer effects,
which broaden the lines in  a manner 
that is not yet satisfyingly represented.

Measurements of the interplanetary Ly $ \alpha $ emission line profile
obtained with a hydrogen absorption cell on board Prognoz 5/6
[{\it Lallement et al.}, 1984; {\it Bertaux et al.}, 1985] have yielded
a good estimate of the bulk velocity ($ 20 \pm 1$ km s$^{-1}$). In this case,
$\mu$  was determined by adjustment of the model. 
However, the \textquotedblleft hot" model never reproduced the data 
in all directions simultaneously. 
More recently, Ly $\alpha$ spectral measurements 
[{\it Lallement et al.}, 1996,
{\it Clarke et al.}, 1998]
confirmed that the inflow speed of H atoms far from the Sun (at the TS)
is within the range of 18 - 21 km s$^{-1}$ using estimates of $\mu$ based on
solar Ly $\alpha$ measurements corresponding to the periods of observation.

According to the results of our calculations (Figure 4b),  
the velocity interval 
18 km s$^{-1} < V_{\rm H,TS} < 21$ km s$^{-1} $ 
corresponds
to the range
 \begin{equation}
 0.07 \: {\rm cm}^{-3} < n_{\rm p, LIC} 
 < 0.2 \: {\rm cm}^{-3}
\end{equation}

The analysis of the Prognoz 5/6  H cell data also provided the \textquotedblleft line profile" 
temperature (8000 $\pm$ 1000 K)
of the H atoms in the inner heliosphere. This is about 1500-2500 K above 
the helium temperature. However, it is important to note that only 
line-profiles observed in the direction perpendicular to the main flow could be 
used and traced back through modeling to the temperature \textquotedblleft at infinity," 
i.e., before the interaction with the Sun (but still inside the heliospheric interface). 
In addition to the uncertainties on the effect of radiative transfer on line profiles, 
this determination may have been slightly biased by the assumption of a given 
temperature in the absorption cell and the temperature slightly underestimated.
Now, the Hubble Space Telescope Ly $\alpha$ spectral observations mentioned above have shown that the line profiles 
on the upwind and downwind side are much larger
than what predict classical models for a gas initially at the same temperature 
as helium. Inferred line-of-sight temperatures are as high as 15,000-20,000 K 
[{\it Clarke et al.}, 1998]. This is in favor of a nonnegligible heating of hydrogen 
at the interface. Still, the role of radiative transfer has to be assessed before 
one can derive a confidence interval for the kinetic temperature.

\subsection{Heliospheric Radio Emissions}
The radio emission detected by Voyager 1 and 2 spacecraft is another type
of heliospheric interface diagnostics.
It is believed that the emission region is connected
with the vicinity of the heliopause.
Major events of kilohertz emission observed in 1983-1984 and in 1992-1993 were
associated with intense solar wind solar events in 1982 and 1991.
The propagation delay in both cases is about 400 days.
Using measurements of the propagation speed of the
interplanetary shocks produced by these events and the time delay of the
onset of the radio burst,
{\it Gurnett and Kurth} [1996] could estimate the distance to the heliopause 
and found that it ranges from 110 to 160 AU.
The comparison with our calculations (Figure 2)
shows that this range for the HP location (for the upwind side) corresponds
to an interstellar proton number density
\begin{equation}
0.08 \: \textnormal{cm}^{-3} < n_{\rm p, LIC} < 0.22  \: \textnormal{cm}^{-3}
\end{equation}
The second feature is the lower frequency (1.8-2.1 kHz) emission band and
 in particular the well-defined \textquotedblleft cutoff" 
of this emission at 1.8 kHz.
This sharp \textquotedblleft cutoff" could be related to
the LIC density, which is the unique constant parameter for the whole interface. 
The plasma frequency at 1.8 kHz implies
an interstellar electron density of  0.04 cm$^{-3} $. 
As explained in the previous section, any radiation emitted in the interstellar medium is, 
in principle, prevented to enter the interface due to a maximum 
compression region characterized by a density we have represented in Figure 5.
Despite these difficulties, it remains that there is no other explanation
for the cutoff, and the particular value $n_{\rm p, LIC}= 0.04$ cm$^{-3} $ 
has a high probability of being the true circumsolar interstellar density.

\subsection{A Synthesis?}
It can be seen from equations (6) and (7) that there is already a small discrepancy between
the interstellar plasma densities obtained from Ly $\alpha$ intensity measurements
and the Ly $\alpha$ profile (Doppler's shift) measurements.
At the same time, both intervals have an intersection 
with the interval in equation (5)
derived from pick-up ions measurements.
This discrepancy disappears if both the interstellar proton and
neutral  number densities are higher than the values we have assumed. 
In this case, it is possible to have simultaneously a smaller bulk velocity 
and a higher neutral H density in the heliosphere.
However, there is a model-independent limit on the LIC  H atom density 
from
the relative H/He abundance $ n_{\rm H,LIC} < 0.23$ cm$^{-3} $ 
(see section 1). This implies that only a small increase of 
$ n_{\rm H,LIC}$
(and then of $ n_{\rm p,LIC}$) is relevant.
For the maximum value $n_{\rm H,LIC}=0.23$ cm$^{-3}$ and 
$ n_{{\rm p,LIC}}=0.07$ cm$^{-3} $ 
(in agreement with equation (5)), we find $n_{\rm H,TS}= 0.11$ cm$^{-3} $,
assuming the same filtration factor as for
$n_{\rm H,LIC}=0.20$ cm$^{-3}$ and $ n_{\rm p,LIC}=0.07$ cm$^{-3} $.
This value corresponds to
lower limit derived from Ly $\alpha$ intensity measurements.

Up to now, we can conclude
that the pair of values $n_{\rm H, LIC}=0.23$ cm$^{-3} $ and
  $n_{\rm p, LIC}=0.07$ cm$^{-3} $ is in agreement with both
 observations of pick-up ions and Ly $\alpha$ radiation.
Now, we note that
 $ n_{\rm p, LIC} = 0.07$ cm$^{-3} $ is very close to the lower limit in equation (8).
Considering that the source of the 3 kHz emission may not be exactly at the HP
but in the region between the TS and the HP could help to reconcile the
two results. This could be due to, for example,
the influence of the electron impact ionization
on the plasma flow in the region between the TS and the HP.
As a matter of fact, this process leads to a strong plasma density gradient 
between the HP and the TS.

However, in any case, the above values are not compatible
with the 1.8 kHz cutoff, if it corresponds to the interstellar emission.
If we assume now that this is really the case, then we have to explain 
why equations (6) and (8) are not justified.

We can give two possible explanations for the discrepancy with equation (8).
First, it is possible that the source of the emission is not exactly at
the HP but in the region between the TS and the HP, as it was mentioned
above. A second possible explanation is that there is
some additional pressure in the interstellar medium.
It may be a magnetic field pressure [{\it Myasnikov}, 1997; {\it Linde et al.}, 1998; 
{\it Pogorelov and Matsuda}, 1998] 
or a low energy 
cosmic ray pressure [{\it Izmodenov}, 1997; {\it Myasnikov et al.}, 1997]. Owing to this
additional pressure, the HP
would be closer to the Sun as compared with what the present model 
implies for  $n_{\rm p, LIC}$ as low as 0.04 cm$^{-3} $.

The calculation of the interstellar magnetic field (IMF) strength required 
to push
the HP as close as 110-160 AU requires a full MHD model coupled to a 
neutral flow model.  Such a model is not yet available. 
The magnetic field is taken into account in some gasdynamical 
models by modifying the equation of state for the plasma [e.g., {\it Gayley et al.}, 1997], 
which is appropriate in some specific conditions of orientation 
and Mach numbers. 
MHD models have been built but without inclusion of the coupling 
to the neutral  flow. 
Hereafter, we estimate
 the IMF strength
needed to reconcile  $n_{\rm H,LIC}=0.04$ cm$^{-3}$,
connected with the 1.8 kHz cutoff, and HP distance measurements 
by {\it Gurnett and Kurth} [1996].
We use
the following formula deduced from our calculations and Figure 2 in the absence of magnetic and cosmic rays pressure:

\begin{equation}
R_{\rm HP} = 24( 3.6 - P_{\rm LIC} \cdot 10^{12} ) + 110 
\end{equation}
Here $ R_{\rm HP} $ is the distance to the heliopause (upwind direction)
in astronomical units, $P_{\rm LIC} $ is the interstellar pressure
in  dyn cm$^{-2}$ deduced from our parameters. We then replace $P_{\rm LIC} $ by the
more general term
\begin{eqnarray}
 P_{\rm LIC} = n_{\rm p, LIC} \cdot 
( 2 k_b T _{\rm LIC} + m_{\rm H} V^2_{\rm LIC} ) \\
+ \alpha \frac{B^2_{\rm LIC}}{8 \pi} + 
P_{\rm GCR,LIC} \nonumber 
\end{eqnarray}
where $n_{\rm p,LIC}$, $T_{\rm LIC} $,and  $V_{\rm LIC}$ are interstellar proton number
density, temperature, and velocity, respectively, $ B_{\rm LIC} $ is the interstellar 
magnetic field strength, $ \alpha $ is an amplification factor
($ \alpha \approx 0-4 $) determined by the angle between 
the interstellar magnetic field and velocity vector 
[{\it Holtzer}, 1989; see, also,  {\it Frisch}, 1993], and
$ P_{\rm GCR,LIC} $ is galactic cosmic ray pressure.

By doing so, we assume that an additional nongasdynamical pressure
acts on the shape of the heliopause on the same way
as a gasdynamical (proton and electron) pressure.
The influence of interstellar atoms is included 
in the numerical coefficients of equation (9),
 which implies this formula is valid 
only if $ n_{\rm H,LIC}$ is of the order of 0.20 cm$^{-3} $.

It is interesting to note that if
the direction of the interstellar magnetic field is the same
as  the wind direction (in principle the unique possibility
for a two-dimensioned modeling), the heliocentric distance of the
heliopause increases as a result of the magnetic field tension
[{\it Baranov and Zaitsev}, 1995] along the wind axis, whereas this distance
decreases in the wing due to magnetic field pressure.
In this case, the formula (9) does not work.
However, the parallel IMF is probably unlikely because, to reconcile
data, we need the heliopause closer to the Sun.


Our formula (9) probably gives better estimates 
of the HP distance variation as a function of the interstellar pressure
than estimates based on balance between the solar
and interstellar pressures [see, e.g.,  {\it Holtzer}, 1989],
because we take into account the plasma 
compressibility and the influence of interstellar neutrals
on the plasma structure.
Actually, owing to these effects, the heliospheric interface
is a kind of damping region for any changes in the interstellar
pressure.  Using balance pressure only,
one overestimates the variation of the HP distance
due to interstellar pressure.

If we assume that $n_{\rm p,LIC} = 0.04$ cm$^{-3}$ 
and that the GCR pressure is $ P_{\rm GCR} = 0.3 \cdot 10^{-12} $ dyn cm$^{-2}$ 
[{\it Holtzer}, 1989],
we find from equation (9) that the interval  110 AU $< R_{\rm HP} <$ 160 AU  
[{\it Gurnett and Kurth}, 1996] corresponds to  
\[
0.737 \cdot 10^{-12} \: {\rm dyn \: cm^{-2}} 
< \alpha \frac{B^2}{8 \pi} < 2.82 \cdot 10^{-12} \: {\rm dyn \: cm^{-2}} 
\]
For $\alpha$=4 (perpendicular magnetic field) [{\it Holtzer}, 1989],
\[
 2.1 \: \mu {\rm G} < B < 4 \: \mu {\rm G}.
\]
This interval is in agreement with the current
estimates of the interstellar magnetic field strength 
[e.g., {\it Frisch}, 1995]. 
However, it is necessary to note that if indeed the IMF strength is above 
2.1 $\mu$G, then a full \textquotedblleft subalfvenic" model is 
required. 
However, in the subalfvenic case, there is no bow shock and then, to date, no explanation for the 2 kHz radio emission. 

It remains that despite the addition of such a pressure, 
the discrepancy with the bulk velocity measurements equation (6) still remains. 
If $n_{\rm p, LIC} = 0.04$ cm$^{-3} $, then the bulk velocity is
$V_{\rm H, TS} = 22.5$ km s$^{-1} $ (23 km s$^{-1} $ for an interstellar velocity of 25.5 km s$^{-1} $).
As we already discussed, the Doppler shift measurements are sensitive to 
the Ly $\alpha$ radiation pressure. The above value of the bulk velocity 
implies that the radiation pressure is above what has been inferred from the
H cell data (there is less deceleration induced by the interface 
and more due to the radiation pressure).

All the above estimates have been done by implicitely assuming that the additional pressure due to the fraction of neutral gas coupled to the plasma remains of the same order with and without interstellar magnetic field.
Of course, both the intensity and the direction of the magnetic field change the plasma pressure between the bow shock and the heliopause, but they also change the thickness of this region. 
An increase of the plasma compression in all cases corresponds to a decrease of the thickness; That is, the two phenomena tend to compensate, as it can be seen in the results of {\it Linde et al.} [1998] 
and {\it Baranov and Zaitsev} [1995], 
for direction and intensity changes, respectively. 
The filtering depends on the product of plasma density and thickness, 
and, as a consequence, it should not
change dramatically. More accurate computations on the basis of MHD models are needed to quantify such changes.

\section{Conclusions}

We have performed a parametric study which shows how sensitive and 
compatible are the various types of diagnostics of the interstellar plasma density 
$ n_{\rm p, LIC}$, i.e., the interstellar neutrals temperature, number density and 
velocity,
and the radio emissions time delays and frequency ranges.
For the neutrals, there are two regimes: For low values ($ n_{\rm p, LIC}  <0.05$ cm$^{-3} $),
the most sensitive parameters are the neutral H density 
and temperature in the inner heliosphere, while for higher values, 
the H bulk velocity only remains sensitive.

In the light of this study, we have discussed the observational results.
Our main conclusion is that it is impossible to reconcile the
results obtained from all types of data as they stand now. 
There is a need for some modifications of the interpretations 
or the confidence intervals.
Two types of solutions (which are mutually exclusive) seem to be favored:
(1) It is possible to reconcile the pick-up ions and Ly $\alpha$  measurements
with the radio emission time delays if a small additional interstellar 
(magnetic or low energy cosmic ray) pressure is added to the main plasma pressure. 
In this case, $ n_{\rm p, LIC}= 0.07$ cm$^{-3} $ and $ n_{\rm H, LIC}= 0.23$ cm$^{-3} $ is the
favored pair of interstellar densities. However, in this case, 
the low frequency cutoff at 
1.8 kHz of course cannot be connected to the interstellar plasma density, and 
one has to search for another explanation.
(2) The low frequency cutoff at 1.8 kHz is connected to the interstellar plasma density, 
i.e., $ n_{\rm p,LIC}=0.04$ cm$^{-3} $. In this case, the
bulk velocity deduced from Ly $\alpha$ spectral measurement 
is underestimated by about 30-50\%  
(the deceleration is by 3 km s$^{-1}$ instead of 5-6 km s$^{-1}$). 
Model limitations (as the use of a stationary classical hot model to derive 
the bulk velocity [{\it Rucinski and  Bzowski}, 1996]) 
or the influence of a strong solar 
Ly $\alpha$ radiation pressure may play a role.
However, in this case, there is a need for  a significant additional
interstellar pressure as compared with case (1).
If the source of this extra-pressure term is a perpendicular magnetic field,
its strength should be in the interval 
 2.1 $\mu$G $< B <$ 4 $\mu$G, a value in agreement 
with local IMF estimates.
How such an additional field will modify our conclusions on the interstellar plasma density is still an open question. 
However, we do not expect substantial changes, as we have discussed in the section 4.

A need for an additional pressure is in agreement with the conclusions of 
{\it Gayley et al.} [1997] from the analysis of the H wall absorption toward alpha Centauri. However, it remains that since the best model of these authors corresponds to a neutral H density of 0.025 cm$^{-3}$ in the inner heliosphere, at least 4 times smaller than the density derived from the pick-up ions, additional calculations for more realistic densities are still needed.

New observations of the heliospheric gas and ions 
are expected within the next years. In particular, the SWAN instrument on 
board SOHO has gathered a considerable amount of Ly $\alpha$ data. 
Their analysis is in progress and will provide extremely 
precise measurements of 
line-of-sight temperature and bulk velocities of atomic H in all directions, 
as their variations change year after year. This should help to disentangle 
the differences between the interpretations discussed here. In parallel, 
there is a crucial need for MHD model developments and for an unambiguous 
interpretation of the 1.8 kHz cutoff.

\acknowledgements
This work has been done in the frame of the INTAS cooperative project:
\textquotedblleft The Heliosphere in the Local Interstellar Cloud"
and partly supported by the International
Space Science Institute (ISSI) in Bern.
V.B., V.I., and Y.M. have been supported
by the Russian Foundation of Basic Research 
under Grants 98-01-00955
and 98-02-16759.
V.I. has been also supported by the ISSI in Bern and by the MENESR (France).    

Janet G. Luhmann thanks Priscilla C. Frisch and Timur J. Linde
for their assistance in evaluating this paper.

\clearpage

\end{article}

\end{document}

%% file: update.tex
\makeatletter
\let\reset@font\empty
\makeatother